\documentclass[useAMS,usenatbib,usegraphicx]{mn2e}

\title[Precision Astrometry of GD\,66]{Precision Astrometry of the Exoplanet Host Candidate GD\,66}

\author[J. Farihi et al.]{J. Farihi$^1$, J. P. Subasavage$^2$, E. P. Nelan$^3$, H. C. Harris$^2$, C. C. Dahn$^2$,
\newauthor J. Nordhaus$^{4,6}$, D. S. Spiegel$^5$\\
$^1$Department of Physics \& Astronomy, University of Leicester, Leicester LE1 7RH, UK; jf123@star.le.ac.uk\\
$^2$United States Naval Observatory, Flagstaff, AZ 86002, USA\\
$^3$Space Telescope Science Institute, Baltimore, MD 21218, USA\\
$^4$Center for Computational Relativity and Gravitation, Rochester Institute of Technology, Rochester, NY 14623, USA\\
$^5$Institute for Advanced Study, Princeton, NJ 08540, USA\\
$^6$NSF Astronomy and Astrophysics Postdoctoral Fellow}

\begin{document}

\date{}

\maketitle

\label{firstpage}

\begin{abstract}

The potential existence of a giant planet orbiting within a few AU of a stellar remnant has profound implications 
for both the survival and possible regeneration of planets during post-main sequence stellar evolution.  This paper 
reports {\em Hubble Space Telescope} Fine Guidance Sensor and U.S. Naval Observatory relative astrometry of 
GD\,66, a white dwarf thought to harbor a giant planet between 2 and 3\,AU based on stellar pulsation arrival times.
Combined with existing infrared data, the precision measurements here rule out all stellar-mass and brown dwarf
companions, implying that only a planet remains plausible, if orbital motion is indeed the cause of the variations in 
pulsation timing.

\end{abstract}


\section{INTRODUCTION}

Observations of extrasolar planetary systems in the post-main sequence have implications for the survival of 
Earth and the terrestrial planets, and indirectly measure the robustness of planet formation processes.  Precision 
radial velocity surveys are discovering a growing number of giant exoplanets orbiting post-main sequence stars, 
where the hosts are either subgiant \citep{jon11,bow10} or first ascent giant stars \citep{get12,nie09,dol09,sat08,
lov07}.  These systems have undergone relatively mild evolution compared to later stages, and their planet 
properties may only reflect formation processes \citep{cur09}.  However, it appears the effect of star-planet 
tides with increasing stellar radius is non-negligible at this early stage \citep{llo11,han10,vil09}, and eventually 
becomes crucial \citep{nor10}.

Candidate planets orbiting post-red giant branch stars are observed with varying degrees of observational 
evidence \citep{cha11,set10,sil07}, but these few detections currently lack the benefit of independent confirmation 
(e.g., transits) and statistics that have corroborated the conventional exoplanet population.  Planets orbiting these 
more highly evolved stars and stellar remnants provide tests of primordial planet formation, long-term star-planet 
and planet-disk evolution, and second-generation planet formation \citep{wic10,mel09}.  The latter scenario is the 
only viable mechanism to produce the pulsar planets \citep{han09}, and they remain the only post-main sequence 
planets confirmed by independent means (gravitational perturbations; \citealt{wol94,ras92}).  

Planetary systems around white dwarfs offer a glimpse into possible futures of the Solar System 
\citep{ver12,dun98,sac93} and the opportunity to study the composition of planetary solids \citep{zuc10,far09,
jur08}.  Furthermore, and perhaps surprisingly, because white dwarfs outnumber A and F-type stars in the solar 
neighborhood\footnote{http://www.recons.org}, they may represent the majority of the nearest planetary systems 
formed at intermediate-mass stars(!).  While there have been several ground- and space-based searches for giant 
planets around white dwarfs \citep{hog09,far08,mul07,deb05}, to date the only published candidate comes from 
variations in pulsation timing of GD\,66 \citep{mul08}.

This paper describes interferometric and astrometric constraints on stellar and low-mass companions to GD\,66
(WD\,0517$+$307).  The motivation for the study is described in \S2, where the observational data supporting a 
companion are reviewed, and theoretical considerations for planet survival are explored.  The {\em Hubble Space 
Telescope (HST)} Fine Guidance Sensor (FGS) and U.S. Naval Observatory (USNO) observations are described 
in \S3, and companion mass limits are derived from the analysis of these data combined with prior studies.

\section{MOTIVATION}

There are two lines of reasoning that led to a search for stellar-mass companions around GD\,66, and each are 
discussed in turn below.  The first is the continued, increasing trend in the observed pulsation arrival times, and 
the second is the issue of planet survival within a few AU during the post-main sequence evolutionary phases of 
an intermediate-mass stellar host.

\subsection{Observational Considerations}

\citet{mul08} performed photometric monitoring of 15 ZZ\,Ceti stars over the course of four years and identified a 
low-amplitude, sinusoidal variation in the expected arrival times of pulsations in GD\,66.  A single turnover in the
observed minus calculated (or $O-C$) diagram was observed around epoch 2005.3.  Assuming the pulsation 
frequency is perfectly stable, and that deviations in transit time are caused by orbital motion, the magnitude of 
this variation is given by

\smallskip
\begin{equation}
\tau = \frac{am\sin{i}}{cM}
\label{eqn1}
\end{equation}

\smallskip
\noindent
where $M$ and $m$ are the mass of the star and companion respectively, $a$ is the semimajor axis, and $i$ is 
the orbital inclination.  This can be rewritten as function of the orbital period $p$ using Kepler's third law.

\smallskip
\begin{equation}
m\sin{i} = \tau \left(\frac{4\pi^2c^3M^3}{G(M+m)p^2}\right)^{1/3}
\label{eqn2}
\end{equation}

\smallskip
\noindent
Thus, for a given periodicity in the pulsation arrival times, the amplitude is a linear function of companion mass for 
$m\ll M$. 

A sinusoidal fit to the GD\,66 timing data was found with parameters $\tau=3.8$\,s, $p=4.5$\,yr \citep{mul08}, 
thus implying $a=2.4$\,AU and $m\sin{i}=2.2\,M_{\rm Jup}$ for $M=0.64\,M_{\odot}$ \citep{ber04}.  This solution 
predicted a turnover in the $O-C$ diagram in late 2007 that did not occur.  Rather, \citet{mul09} later reported that 
the arrival times continued to increase and a revised fit yielded $\tau\approx5$\,s, $p=5.7$\,yr, and hence $a=2.7
$\,AU, $m\sin{i}=2.4\,M_{\rm Jup}$.  This newer fit predicted a turnover in 2008 that again did not occur \citep{her10}.

Based on this trend of increasing $\tau$, it became possible that the observed signal could be due to a stellar 
companion.  In this case, the initial turnover discovered in the $O-C$ diagram would be a relatively low probability 
event.  Despite being somewhat unlikely, this alternative merits investigation due to the profound implications of a 
planet orbiting a stellar remnant.  Stellar mass solutions to Equation \ref{eqn2} become plausible for $\tau>10$\,s 
and $i>85\degr$, but existing infrared data (see \S3.3) constrain such companions to be degenerate: brown dwarf, 
white dwarf, or neutron star.  Notably, a second white dwarf can remain hidden in optical spectroscopy if at least of 
comparable mass to the primary.  An excellent example of this is PG\,0901$+$140 \citep{far05}, a $3\farcs6$ DA5$
+$DA6 binary that exhibits an apparently-single DA5.5 spectrum \citep{lie05}.  For such a companion mass to be 
viable, the increasing $O-C$ trend observed by \citep{mul09} would have to continue for several years.  Interestingly, 
orbital separations of a few to several AU have been found for five double white dwarf systems using {\em HST} FGS 
observations \citep{sub09,nel07}, lending credibility to a binary scenario.

\begin{table}
\begin{center}
\caption{USNO Observations of GD\,66 over 10.25\,yr}\label{tbl1}
\begin{tabular}{@{}ll@{}}
\hline

Astrometry:&\\
$\pi_{\rm rel}$			&$16.86\pm0.16$\,mas\\
$\pi_{\rm ref}$			&$0.97\pm0.12$\,mas\\
$\pi_{\rm abs}$			&$17.83\pm0.20$\,mas\\

$\mu_\alpha$			&$+55.3\pm0.1$\,mas\,yr$^{-1}$\\
$\mu_\delta$			&$-120.3\pm0.1$\,mas\,yr$^{-1}$\\

$v_{\rm tan}$			&$35.2\pm0.4$\,km\,s$^{-1}$\\

\hline
Photometry:&\\
$V$					&$15.56\pm0.02$\,mag\\
$B-V$				&$+0.13\pm0.02$\,mag\\
$V-I$				&$-0.05\pm0.02$\,mag\\
$M_V$				&$11.82\pm0.02$\,mag\\
\hline

\end{tabular}
\end{center}

{\em Note}.  Photometry is on the Johnson-Cousins system and given in Vega magnitudes.  Conservative errors 
for these mean magnitudes are 0.02\,mag.  

\end{table}

\subsection{Theoretical Considerations}

Because GD\,66 is a carbon-oxygen core white dwarf, it has passed through both the first ascent (RGB) and 
asymptotic giant branch (AGB) phases.  \citet{ber04} give spectroscopically derived stellar parameters for GD\,66 
of $T_{\rm eff}=11\,980\pm200$\,K, $\log\,g\,({\rm cm\,s}^{-2})=8.05\pm0.05$, implying $M=0.64\pm0.03\,M_{\odot}$.  
Using the average of three different initial-to-final mass relations \citep{wil09,kal08,dob06} yields a possible range of 
main-sequence progenitor masses between 2.2 and $2.6\,M_{\odot}$.  As a basic line of reasoning that favors a stellar 
companion over a planet at a few AU, consider that the maximum AGB radius for this range of progenitor stars is in the 
vicinity of 3\,AU \citep{vil07}.  

However, as explored in detail by \citet{nor10} there are at least three processes that determine the orbital fate 
of low-mass companions to post-main sequence stars such as white dwarfs: 1) orbital expansion due to mass loss 
\citep{jea24}, 2) tidal dissipation of orbital energy, and 3) destruction or survival in a common envelope phase.  It 
is well known that very low-mass stellar and brown dwarf companions are capable of surviving within a common 
envelope \citep{max06,far05}  that results from unstable mass transfer from a giant primary \citep{pac76}, and these 
short-period systems are manifest among white dwarf binaries \citep{sch03,sch96}.  The inward pull of tidal torques 
and common envelope evolution, together with the outward expansion of orbits beyond the grasp of frictional and
 tidal forces, effectively create a depopulated region of intermediate orbital separations \citep{nor10}.  This bimodal 
distribution of low-mass, unevolved companions to white dwarfs, barren from $\sim0.1$ to $\sim10$\,AU, has been 
empirically confirmed using high-resolution optical imaging with {\em HST} \citep{far10,far06} and by spectroscopic 
and photometric monitoring from the ground \citep{neb11}.

Following the prescription of \citet{nor10}, the relevant semimajor axis boundaries for single planets at GD\,66 were 
computed including mass loss, tidal forces, and common envelope evolution.  The main-sequence progenitor masses 
span the range of values expected for this system (see \S3.1), while companion masses include the $m\sin{i}=2.4\,M_
{\rm Jup}$ determined from the pulsation timing analysis, and a 7\,$M_{\rm Jup}$ upper limit estimated from infrared 
photometry and modeling (see \S3.3; \citealt{mul09}).  Included in these calculations are the largest, initial semimajor 
axis for planets directly engulfed by the AGB star, and smallest, final semimajor axis for planets that avoid being 
swallowed.  Planets are destroyed in the first instance, while in the latter case their orbits are strongly influenced by 
tides yet just avoid the giant envelope.  It is found that for a wide range of possible tidal prescriptions, all companions 
that avoid engulfment end up in $a\geq3.6$\,AU orbits, and all but the most finely tuned initial conditions lead to final 
separations several AU larger.  Therefore, an extant planet at $2-3$\,AU around a white dwarf remnant of a $M\geq2.2
\,M_{\odot}$ main-sequence star would require an unexpected evolutionary scenario (e.g., capture or re-formation).
These predictions imply that a planet is rather unlikely, and largely prompted a search for stellar-mass objects capable 
of producing the sinusoidal timing variations observed by \citep{mul08}.

\begin{figure}
\includegraphics[height=86mm,angle=90]{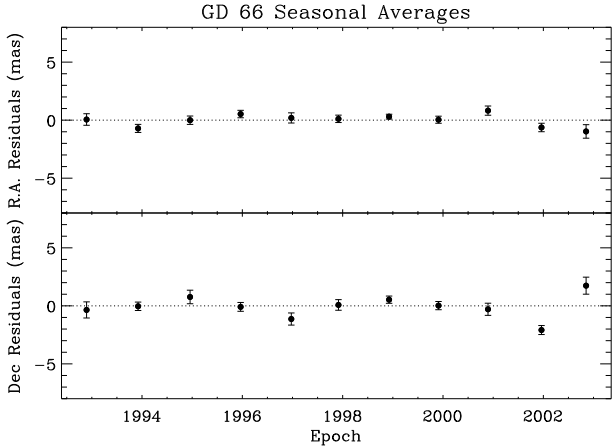}
\caption{Residuals obtained after fitting proper motion and parallax to the USNO astrometric observations of GD\,66.
Shown are the seasonal averages and the standard error of the mean.  A total of 172 frames taken on 151 nights were 
used in the parallax solution, with 9 to 32 observations per season.  The behavior of the residuals in Declination are also 
observed in a field star near to GD\,66 on the CCD, and are thus systematic.
\label{fig1}}
\end{figure}

\section{OBSERVATIONS AND DATA}

\subsection{USNO Astrometry}

Optical CCD observations of GD\,66 spanning just over a decade were carried out as part of the USNO faint star 
parallax program.  Imaging photometry was collected for the purposes of correcting for differential color refraction and 
to measure an absolute trigonometric parallax.  In practice, because observations were taken within (and usually much 
less than) one hour of meridian, differential color refraction was minimal.  A complete discussion of the astrometric data 
acquisition and reduction can be found in \citet{dah02}.  Briefly, the observations were taken using the USNO 1.55\,m 
Strand Astrometric Reflector and Tektronix $2048\times2048$ camera with 24.0\,$\mu$m pixels at $0\farcs325$\,pixel$
^{-1}$.  These observations employed a wide $R$-band filter, similar to that described in \citet{mon92}.  Independent 
photometry in the Johnson $BV$ and Cousins $I$ band was collected on two nights at the USNO 40\,in telescope.  
Photometric standards of various colors from \citet{lan92} were taken at multiple airmasses to correct for extinction 
and color terms.

\begin{figure}
\includegraphics[width=86mm,angle=0]{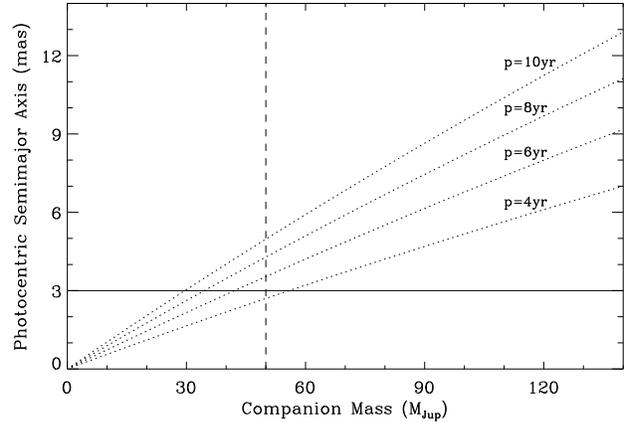}
\caption{Companion mass sensitivity for primary mass 0.66\,$M_{\odot}$ and a $3\sigma$ threshold of 3\,mas deviation 
due to circular orbital motion for GD\,66.  Plotted as dotted lines are the astrometric excursion as a function of companion 
mass for several benchmark orbital periods.  The dashed line corresponds to a detectable companion mass of 50\,$M_{\rm 
Jup}$ at nearly all orbital periods between 4 and 20\,yr.  For circular motion, the companion mass sensitivity is immune to 
orbital inclination, and thus the USNO astrometry rule out a wide range of low-mass, stellar and substellar companions 
capable of producing the $O-C$ variations.
\label{fig2}}
\end{figure}

\begin{figure*}
\includegraphics[width=178mm]{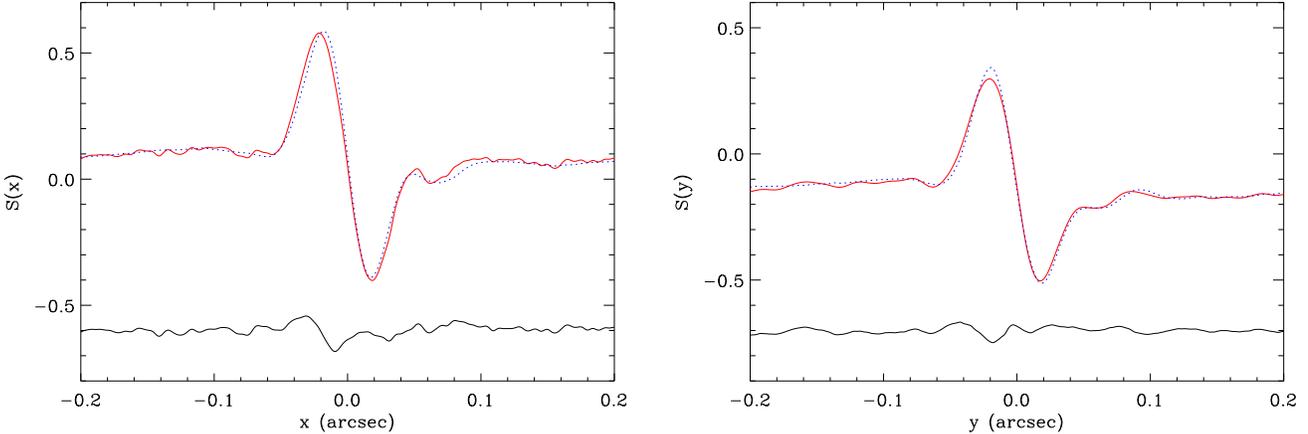}
\caption{Interference fringes for GD\,66 along the FGS1r x- and y-axes.  The fringes of the science target are shown 
as solid red lines and do not differ significantly from that of the calibration point source, BD$+$84\,12, shown as dotted 
blue lines.  At the bottom of each panel are shown the residuals in black, which arise from photometric noise, indicating 
that GD\,66 is unresolved.  These data rule out stellar companions at $\Delta m=2.44$\,mag in F538W with separations 
greater than 10\,mas, or 0.56\,AU at the trigonometric parallax distance.
\label{fig3}}
\end{figure*}

The USNO astrometric results are listed in Table \ref{tbl1}, placing GD\,66 at $d=56.1^{+0.6}_{-0.5}$\,pc, and yielding 
an absolute visual magnitude of $M_V=11.82$\,mag.  Comparing this with the spectroscopically derived value of $M_V
=11.75$\,mag based on $T_{\rm eff}=11\,980$\,K \citep{ber04}, the parallax favors a slightly higher mass.  Using this
same effective temperature, white dwarf atmospheric models yield $\log\,g=8.09$, $M=0.66\,M_{\odot}$, and a cooling 
age of 420\,Myr \citep{fon01}.  Although the astrometric data presented here do not constrain the temperature of GD\,66, 
the the ZZ\,Ceti instability strip is narrow at $\Delta T_{\rm eff}\approx1000$\,K for $\log\,g\approx8.1$ \citep{gia05}.  To
within the errors introduced by its few \% photometric variability \citep{fon85}, the optical $BVI$ photometry of GD\,66 
yields colors precisely as expected for a DA white dwarf of the published effective temperature.

Figure \ref{fig1} plots the astrometric residuals for each observing season after subtracting the proper motion and 
parallax, demonstrating a few mas precision.  This relative astrometry is sensitive to orbital motion induced by an {\em 
optically-dark} companion.  Notwithstanding the systematically noisier residuals in Declination relative to those in Right 
Ascension, the astrometric monitoring of GD\,66 should have detected a $3\sigma$ deviation of 3\,mas due to orbital 
motion in the plane of the sky.  Figure \ref{fig2} marks this detection threshold in a plot of the astrometric excursion as 
a function of companion mass, for several benchmark orbital periods in a circular binary configuration.  These mass
limits are independent of orbital inclination for circular orbits, and degrade only in the specific case of an eccentric 
orbit at high inclination with line of apsides near to the line of sight ($\omega\approx90$\degr).

Because the astrometric companion sensitivity only diminishes for finely-tuned orbital parameters, the USNO data rule
out all $m\ga50\,M_{\rm Jup}$ (0.05\,$\,M_{\odot}$), optically-faint companions, including very low-mass stars, neutron 
stars and black holes, for orbital periods $p>4$\,yr.  Remarkably, a period-dependent range of relatively low, brown 
dwarf companion masses can be similarly ruled out for longer periods up to 20\,yr.  However, these data provide little 
or no constraints on white dwarf companions in this mass range, as a binary of equal mass and brightness will exhibit 
no photocenter shift.

\subsection{FGS Interferometry}

GD\,66 was observed during {\em HST} Cycle 18 on 6 October 2010 by FGS1r in its high angular resolution Transfer mode, 
using the F538W filter which covers $4500-7000$\AA.  In this mode FGS1r repeatedly scans an object and provides data from 
which interference fringes along its two orthogonal axes can be reconstructed \citep{nel11}.  A relatively bright comparison 
star known to be a point source at FGS resolution, BD$+84$\,12 was observed as a calibration source.  Figure \ref{fig1} plots 
the FGS data for GD\,66 and calibration star, and reveals the white dwarf is unresolved to 10\,mas and $\Delta m=2.44$\,mag.

The FGS interferometry is sensitive to white dwarf companions at nearly all possible orbital separations and inclinations.
While two equal mass degenerates would be readily detected down to 0.56\,AU and $p=0.4$\,yr, such short periods are 
not consistent with the $p>4$\,yr, pulsation timing variations of GD\,66.  For the periods fitted to the timing data of GD\,66, 
Equation \ref{eqn1} dictates that a stellar-mass companion should be within several degrees of face-on.  Therefore, the 
FGS data would have readily detected orbital separations from a few to several tens of AU, and thus rule out companions 
of comparable brightness and mass.  However, in the very low probability event that the $O-C$ turnover was a 1 in $10^5
$ chance detection, the FGS data also constrain a putative double degenerate binary for any inclination.  If the observed 
$O-C$ minimum in 2005.3 corresponds to a binary system in conjunction as seen from Earth, then the projected separation 
of the stars has been widening since, yet must still have an angular separation below 10\,mas $=0.56$\,AU from its non-$
$resolution by FGS in 2010.8.  This implies a system period greater than $10^5$\,yr for edge-on orbits and total system 
masses $M\ga1\,M_{\odot}$.  All double white dwarfs with shorter periods would have had a wider projected separation 
and been detected by the FGS.

Importantly, white dwarf secondaries fainter than the primary by $\Delta V > 2.4$\,mag are not feasible by normal, 
unperturbed stellar evolution; the total system age is insufficient to achieve such low luminosities.  From the 0.66\,$M_
{\odot}$ white dwarf mass derived from the trigonometric parallax, the main-sequence progenitor of GD\,66 had a mass 
between 2.4 and 2.8\,$M_{\odot}$ \citep{wil09,kal08,dob06}, a hydrogen-burning lifetime in the range $460-660$\,Myr 
\citep{hur00}, and a total system age of $0.9-1.1$\,Gyr.  Any cool and massive white dwarf (total age $\approx$ cooling 
age) can become no fainter than $M_V=13.9$\,mag over 1.1\,Gyr \citep{fon01} and would be detectable as $\Delta V 
<2.1$\,mag.

\subsection{Infrared Photometry}

Existing {\em Spitzer} IRAC photometry of GD\,66 are sensitive to substellar companions that would not be detected 
by the USNO or FGS optical astrometry.  \citet{mul09} used these infrared observations to estimate an upper limit on 
spatially-unresolved (within $2\farcs4\approx130$\,AU), self-luminous objects of 7\,$M_{\rm Jup}$, thus indicating the 
putative companion should have a planet-sized mass.  Their method utilized observations of three white dwarfs with 
similar temperatures to constrain the photospheric flux ratio between 3.6 and 4.5\,$\mu$m, reporting a $1\sigma$ total 
uncertainty of 0.6\% in the ratio observed for GD\,66.  As an example of an alternative method using IRAC photometry, 
\citet{far08} establish substellar companion mass limits at 15\% above the predicted (or measured) 4.5\,$\mu$m 
photospheric flux.  Applying this procedure to GD\,66 with the USNO parallax distance, a total system age of 1\,Gyr 
(see \S3.2), and assuming its 4.5\,$\mu$m flux measurement \citep{mul09} is due to the photosphere, one obtains an 
upper limit of 11\,$M_{\rm Jup}$ \citep{bar03}.  If one insists on only a 10\% excess at this wavelength, the upper limit 
drops to and 9\,$M_{\rm Jup}$.  It should be mentioned that all such analyses depend on atmospheric models, none 
of which have empirical constraints at these masses, as well as the adopted total system age.  Regardless of the method, 
the infrared data rule out all but planetary masses, according to models.

\section{CONCLUSIONS}

The combined observational data on GD\,66 limit any binary companions orbiting within several AU to planetary masses.  
Specifically, any long-term, increasing trend in the pulsation arrival times cannot be due to stellar-mass secondaries, which 
include low-mass stars, white dwarfs, neutron stars, and black holes with periods longer than 4\,yr.  The USNO relative 
astrometric monitoring of just over a decade rules out stellar-mass, dark companions with periods between 4 and 20\,yr, 
while the FGS observations rule out virtually all white dwarf companions, regardless of orbital inclination.  Based on
substellar cooling models, infrared data further restrict low-mass companions within a hundred AU to have planetary 
masses.  It is noteworthy that the trigonometric parallax and infrared photometry by themselves rule out a range of double 
degenerate scenarios, but the astrometric monitoring and interferometry provide significantly more stringent limits to such 
binaries.

If the observed timing data at GD\,66 are due to orbital motion, these new and exiting data rule out a vast range of realistic 
companion masses, and strengthen the likelihood of a planet-sized mass as the cause.  Because stellar evolution models 
and resulting star-planet interactions indicate a planet within a few AU of an intermediate-mass star is not likely to survive 
to the white dwarf stage, the putative planets at GD\,66 and other post-RGB stars V391\,Peg and KPD\,1943+1405 (KOI\,55), 
if real, may have been dynamically injected or formed in a second-generation of planet formation.  Further study of these 
and future, post-RGB planet candidates is needed to better understand the population and architecture of planetary systems 
around white dwarfs.

\section*{ACKNOWLEDGMENTS}
The authors thank the referee, F. Mullally for a constructive and helpful report.  DSS gratefully acknowledges support 
from NSF grant AST-0807444 and the Keck Fellowship.  JN is supported by an NSF Astronomy and Astrophysics 
Postdoctoral Fellowship under award AST-1102738 and by NASA HST grant AR-12146.04-A.

\label{lastpage}

\end{document}